\documentclass[aps,prb,twocolumn,showpacs]{revtex4}  
\usepackage[T1]{fontenc}
\usepackage[latin1]{inputenc}
\usepackage{amssymb}
\usepackage{epsfig}
\usepackage{longtable}
\makeatletter

\usepackage{dcolumn}
\usepackage{amstext}
\usepackage{graphicx}
\usepackage{bm}

\begin{document}

\title{Spin Decay in a Quantum Dot Coupled to a 
Quantum Point Contact}

\author{Massoud Borhani}

\author{Vitaly N. Golovach}

\author{Daniel Loss}

\affiliation{Department of Physics and Astronomy, 
University of Basel, Klingelbergstrasse 82, 
4056 Basel, Switzerland}

\date{\today}

\begin{abstract}
We consider a mechanism of spin decay for an electron 
spin in a quantum dot due to coupling to a nearby  quantum 
point contact (QPC) with and without an applied bias voltage.
The coupling of spin to charge is induced by  the  
spin-orbit interaction in the presence of a magnetic field.
We perform a microscopic calculation of  the effective Hamiltonian
 coupling constants to obtain the 
QPC-induced spin relaxation and decoherence rates in a realistic system. 
This rate is shown to be proportional to the shot noise
of the QPC in the regime of large bias voltage and scales
as $a^{-6}$ where $a$ is the distance between the quantum dot and the QPC.
We find that, for some specific orientations of the setup
with respect to the crystallographic
axes, the QPC-induced spin 
relaxation and decoherence rates vanish, while the charge sensitivity of
the QPC is not changed. This result can be used in experiments to minimize 
QPC-induced spin decay in read-out schemes.  

\end{abstract}
\maketitle

\section{Introduction}
Recent progress in nanotechnology has enabled access to the 
electron spin in semiconductors in unprecedented ways, 
\cite{ALS,WolfScience,DasSarma} with the electron spin in  quantum dots
being a promising candidate for a qubit due
to the potentially long decoherence time of the spin.\cite{LD,Vero}
 Full understanding of the decoherence processes of the
electron spin is thus crucial. 
On the other hand, as a part of a quantum computer, read-out 
systems play an essential role in
determining the final result of a quantum computation.  
However, read-out devices, in general, affect the  spin state of the
system in an undesired way.
 Quantum point contacts (QPCs) which are used as  charge
detectors,\cite{Buks} in particular, couple to the spin 
 via the spin-orbit interaction. For small GaAs quantum dots, the spin-orbit
length ($\lambda_{SO} \approx 8\;\mu$m) is much larger than the dot size 
($\lambda_{d} \approx 50$ nm) and thus the spin-orbit interaction
presents a small perturbation. Nevertheless, we will see that  
shot noise in the QPC can induce an appreciable spin decay 
via this weak spin-orbit coupling.

Quite remarkably, the number of electrons in quantum
dots can be tuned starting from zero.
\cite{Tarucha,Ciorga,Elzerman,PettaPRL}
More recently, Zeeman levels have been resolved \cite{Hanson} 
and the spin relaxation time ($T_1$) has been measured,
 yielding times of the order of milliseconds in the 
presence of  an in-plane magnetic field  of $8$ T.\cite{ENature,HansonCM}
 In these experiments, based on spin-charge conversion,\cite{LD} 
use is made of  a QPC located near the quantum dot 
as a sensitive charge detector to monitor changes of the
 number of electrons in the  dot.
The shot noise in the QPC affects the electron charge 
in the quantum dot via the Coulomb interaction,\cite{Buks,Field} and
therefore, it can couple to the electron spin as well,
 via the spin-orbit interaction.
While charge relaxation and decoherence in a quantum dot due to a 
nearby functioning QPC have been studied
before,\cite{Levinson,Aleiner} we show here that the same charge
fluctuations in the QPC introduce spin decay via spin-orbit 
and Zeeman interactions.
Note that several read-out schemes utilizing a QPC  have been
considered before\cite{nshot} in the context of the spin qubit. 
However,  in Ref.~\onlinecite{nshot} the QPC was used for charge
read-out, while the spin state of the qubit was converted into 
the charge state of a {\em reference} dot.\cite{LD}
Recently, a different read-out scheme has been 
implemented,\cite{ENature} in which the
reference dot was replaced by a Fermi lead and the QPC was
coupled directly to the spin qubit.

The effect of spin-orbit interaction on spin relaxation and decoherence
was considered in Ref.~\onlinecite{GKL}.
There, it was shown that the decoherence time $T_2$ due to 
spin-orbit interaction approaches its upper bound,\cite{GKL} 
i.e. $T_2=2T_1$, determined by spin-flip processes.\cite{GKL,KhaNaz}
Measurements of $T_1$ have been performed on spins in electrostatically
confined (lateral) quantum dots\cite{ENature} ($T_1\simeq 0.85\,{\rm ms}$) 
and self-assembled quantum dots\cite{Kroutvar} 
($T_1\simeq 20\,{\rm ms}$).
The measured spin relaxation times $T_1$ in both cases
agree well with the theory in Refs.~\onlinecite{GKL} and 
\onlinecite{KhaNaz}.
In addition to the spin-orbit interaction, 
the hyperfine interaction plays an important 
role in quantum dots.
\cite{BLD,KhaetskiiLossGlazman,MerculovEfrosRosen,
ErlingssonNazarov,SchliemannKhaetskiiLossNUCL,
SousaDasSarma,Bill,Bracker,Koppens,PettaT2Nature,PettaScience}
Measurements of the spin decoherence time $T_2$ have recently been
performed in a self-assembled quantum dot\cite{Bracker} 
($T_2^*\simeq 16\,{\rm ns}$) as well as
in a double-dot setup for singlet-triplet decoherence
($T_2\simeq 10\,{\rm \mu s}$).\cite{PettaScience}
Finally we note that a number of alternative schemes to measure the
decoherence time of the electron spin in quantum dots have been 
proposed.\cite{Engel,GLA,Gywat}

Motivated by these recent experiments, 
we study here the effect of the QPC on spin relaxation 
and decoherence in the quantum dot.
For this, we first derive an effective Hamiltonian for the spin
dynamics  in the quantum dot and find a transverse 
(with respect to the external magnetic field) fluctuating magnetic field.
We calculate microscopically the coupling constants of the effective 
Hamiltonian by modeling the QPC as a one-dimensional 
channel with a tunnel barrier.
We show that this read-out system speeds up the spin
decay and derive an expression 
for the spin relaxation time $T_1$. However, there are 
some regimes in which this effect vanishes,
in the first order of spin-orbit interaction. The relaxation time will turn
out to be strongly dependent on the QPC orientation on the substrate, 
the distance between the QPC
and the quantum dot, the direction of the applied magnetic field, the
Zeeman splitting $E_Z$, the QPC transmission coefficient ${\cal T}$,
and the screening length $\lambda_{sc}$ (see Fig. \ref{QPC}).    
Although this effect is, generally, not larger than other 
spin decay mechanisms (e.g. coupling of spin to 
phonons \cite{GKL} or nuclear spins \cite{Bill}), it is
still measurable with the current setups under certain conditions.
The following results could be of interest to experimentalists to minimize 
spin decay induced by QPC-based charge detectors.\\  

The paper is organized as follows. In Section II we introduce our model for 
a quantum dot coupled to a quantum point contact and the corresponding 
Hamiltonian. Section III is devoted to the derivation of the 
effective Hamiltonian  for the electron spin in the quantum dot.
In Section IV we derive  microscopic expressions for 
the coupling constants of  the effective Hamiltonian and discuss different 
regimes of interest.  Finally, in Section V, we calculate the electron spin 
relaxation time $T_1$ due  to the QPC and make  numerical
predictions for typical lateral quantum dots.\\

\section{The Model}
\begin{figure}\vspace{0.cm}\narrowtext
{\epsfxsize=7.5cm
\centerline{{\epsfbox{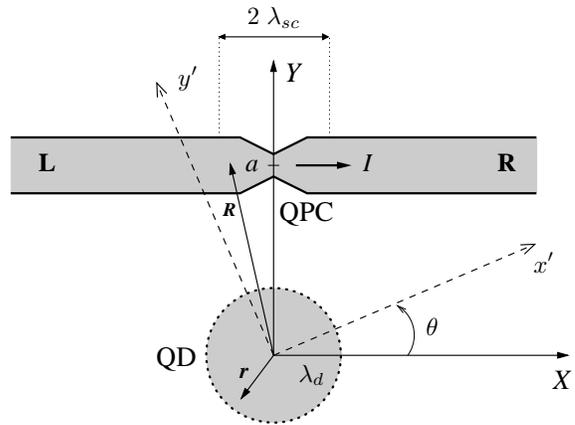}}}}
\caption{\small Schematic of the quantum dot (QD) coupled to a QPC. 
The $(X,Y)$ frame gives  the setup orientation, left (L) and right (R) leads,
with respect to the crystallographic  directions 
$x'\equiv[110]$ and $y'\equiv[\bar110]$.
The dot has a radius $\lambda_d$ and is located at a distance $a$ from
the QPC. The vector ${\bm R}$ describes the QPC electrons and ${\bm r}$ 
refers to the  coordinate of the electron in the dot.
The noise of the QPC current $I$ perturbs the electron spin on the dot
via the spin-orbit interaction.}
\label{QPC}
\end{figure}

We consider an electron in a quantum dot and a nearby functioning 
quantum point contact (QPC),
see Fig.~\ref{QPC}, embedded in a two-dimensional electron gas (2DEG).
 We model the QPC as a one-dimensional wire
coupled via the Coulomb interaction to the electron in the quantum dot. 
We also assume that there is only one electron inside the dot, which
is feasible experimentally.
\cite{Tarucha,Ciorga,Elzerman,PettaPRL,Hanson,ENature}
The Hamiltonian describing this coupled system reads
$H = H_d + H_Z + H_{SO} + H_Q + H_{Qd}$, where 
\begin{eqnarray}
H_d &=& \frac{ p^2}{2m^*} + U(\bm{r}),\\
H_Z &=& \frac{1}{2} g\mu_B{{\bm B}} \cdot {{\bm\sigma}}
 = \frac{1}{2}E_Z {{\bm n}} \cdot {\bm\sigma}\label{HZ},\\
H_{SO}& =& \beta(-p_x\sigma_x + p_y\sigma_y) + \alpha(p_x\sigma_y -
p_y\sigma_x)\label{Hso},\\
H_Q &=& \sum_{lk\sigma} \epsilon_k \bar{C}^\dagger_{lk\sigma} 
\bar{C}_{lk\sigma}\label{HQ},\\
H_{Qd} &=&\sum_{ll'kk'\sigma} \eta_{ll'}({\bm r}) \bar{C}^\dagger_{lk\sigma} 
\bar{C}_{l'k'\sigma}\label{HQd}. 
\end{eqnarray}
Here,  $Q$ refers to the  QPC and
$d$ to the dot, ${\bm p} = -i \hbar {\bm\nabla} + (e/c)\bm{A}(\bm{r})$ is
the electron 2D momentum, $U(\bm{r})$ is the lateral confining
potential, with ${\bm r} = (x,y)$,
$m^*$ is the effective mass of the electron, and $\bm{\sigma}$ are the Pauli
matrices. The 2DEG is perpendicular to the $z$ direction. 
The spin-orbit Hamiltonian $H_{SO}$ in Eq.(\ref{Hso})
includes both Rashba \cite{Rashba} spin-orbit coupling ($\alpha$), 
due to asymmetry of the quantum well
profile in the $z$ direction, and Dresselhaus \cite{Dress} spin-orbit
couplings ($\beta$), due to the  inversion asymmetry of the GaAs lattice.
 The Zeeman interaction $H_Z$ in Eq. (\ref{HZ}) introduces a
spin quantization axis along ${{\bm n}} = {{\bm B}}/B = (\cos
\varphi\sin \vartheta, \sin \varphi \sin \vartheta, \cos \vartheta)$.
The QPC consists of two Fermi liquid leads coupled via a tunnel barrier and
is described by the Hamiltonian $H_Q$, where  
 $\bar{C}_{lk\sigma}^\dagger$, with $l =L, R $, creates an 
electron incident from lead $l$, with wave vector $k$ and spin 
$\sigma$.
We use the overbar on, e.g., $\bar{C}_{lk\sigma}$ to denote
the scattering states in the absence of electron on the dot. 
The Hamiltonian $H_{Qd}$ in Eq.~(\ref{HQd}) describes the
coupling between the quantum dot electron and  
the QPC electrons.
We assume that the coupling is given by the
screened Coulomb interaction, 
\begin{eqnarray}
\eta_{ll'}({\bm r}) &=& \langle \overline{lk}| \frac{e^2}
{\kappa|{\bm r} - {\bm R}|}\tilde{\delta}({\bm R}-{\bm a})|
\overline{l'k'}\rangle,\label{eta}
\end{eqnarray}
where ${\bm R}=(X,Y)$ is the  coordinate of the electron in the QPC 
and $\kappa$ is the dielectric constant.
The Coulomb interaction is modulated by a dimensionless screening factor 
$\tilde{\delta}({\bm R}-{\bm a})$, 
\footnote{Strictly speaking, the screening factor depends also
on ${\bm r}$, $\tilde{\delta}({\bm R}-{\bm a},{\bm r})$.
However, since usually $\lambda_d \lesssim \lambda_{s}$, we approximate 
$\tilde{\delta}({\bm R}-{\bm a},{\bm r})\approx 
\tilde{\delta}({\bm R}-{\bm a},0) \equiv \tilde{\delta}({\bm R}-{\bm a})$,
keeping in mind that $|{\bm r}| \lesssim \lambda_d$.}
where ${\bm a}=(0,a)$ gives the QPC position (see Fig.~\ref{QPC}).
The quantum dot electron interacts with 
 the QPC electrons mostly at the tunnel barrier;
away from the tunnel barrier the interaction is screened
due to a large concentration of electrons in the leads.
For the screening factor we assume, in general, a
 function which is peaked at the QPC and has
a width $2\lambda_{sc}$ (see Fig.~\ref{QPC}).
Note that $\lambda_{sc}$ is generally different from the screening length in
the 2DEG and depends strongly on the QPC geometry and size.
Generally, $\eta_{ll'}$ are  $k$-dependent, however, their $k$-dependence
turns out to be weak and will be discussed later.

\section{The Effective Hamiltonian}
The quantum dot electron spin couples to  charge
fluctuations in the QPC via the spin-orbit Hamiltonian (\ref{Hso}).
The charge fluctuations are  caused by electrons passing
through the QPC.
To derive an effective Hamiltonian for the coupling of spin to charge
fluctuations, we perform a Schrieffer-Wolff transformation,\cite{Mahan} 
$\tilde H = \exp(S) H \exp(-S)$, and remove 
the spin-orbit Hamiltonian in leading order. 
We thus require that $\left[ H_d + H_Z, S \right] =  H_{SO} $,
under the condition $\lambda_d\ll\lambda_{SO}$, where
$\lambda_d$ is the quantum dot size and
$\lambda_{SO} = \hbar/m^* (|\beta|
+ |\alpha|)$ is the minimum spin-orbit length. 
 The transformed Hamiltonian is then given by \\
\begin{eqnarray}
\tilde H &=&  H_d + H_Z + H_Q + H_{Qd}+ \left[ S , H_{Qd}
  \right],\label{Ht}\\
S &=& \frac{1}{L_d+L_Z}H_{SO}
=\frac{1}{L_d}\sum_{m=0}^{\infty} 
\left( -L_Z\frac{1}{L_d}\right)^m H_{SO}\label{S},\;\;\;\;\;\\
H_{SO}&=&iL_d({\bm \sigma}\cdot{\bm \xi}),
\end{eqnarray}
where $L$ is Liouville superoperator for a given Hamiltonian
 defined by $L A \equiv [H,A]$ and 
${\bm \xi}$ is a vector in the 2DEG plane and has a simple form in the
coordinate frame $x'=(x+y)/\sqrt{2}$, $y'=(y-x)/\sqrt{2}$, $z'=z$,
namely, ${\bm \xi}=(y'/\lambda_-, x'/\lambda_+, 0)$, where
$\lambda_{\pm} = \hbar/m^* (\beta \pm \alpha)$ are the spin-orbit 
lengths. For a harmonic dot confinement 
$U(r)=\frac{1}{2}m^*\omega_0^2r^2$, we have
\begin{eqnarray}
\frac{1}{L_d}x &=& \frac{-i}{\hbar m^*\omega_0^2}
\left( p_x + \frac{e B_z}{c} y \right), \label{L1}\\
\frac{1}{L_d}y &=& \frac{-i}{\hbar m^*\omega_0^2}
\left( p_y - \frac{e B_z}{c} x \right),\label{L2}\\
\frac{1}{L_d}p_j &=& \frac{im^*}{\hbar}r_j,
\;\;\;\;\;\;\;\;\;\;\;\;\;(j= x,y).\label{L3}
\end{eqnarray} 
 In addition, we have the following relations for the Zeeman
Liouvillian
\begin{equation}
L^m_Z({\bm \sigma}\cdot{\bm \xi}) =
\left\{\begin{array}{l}
\;\; i E_Z^m [{\bm n}\times{\bm \xi}]\cdot{\bm \sigma},
\;\;\;\;\;\;\;\;\; \mbox{for odd }  m > 0 \\
-E_Z^m[{\bm n}\times({\bm n}\times{\bm \xi})]\cdot{\bm \sigma},
\;\;\; \mbox{for even }  m > 0,\;\;\;\;\;\label{LZ} 
\end{array}\right.
\end{equation}
where $E_Z = g \mu_B B $ is the Zeeman splitting. 
The last term in Eq. (\ref{Ht}) gives the coupling of the dot spin to
the QPC charge  fluctuations. The transformation matrix $S$ (to
first order in spin-orbit interaction) can be derived 
by using the above relations (see  Appendix A). We obtain
\begin{eqnarray} 
-i S &=& 
\mbox{\boldmath$\xi$}\cdot\mbox{\boldmath$\sigma$}+ \left[\mbox{\boldmath$n$}
\times \mbox{\boldmath$\xi$}_1\right]\cdot\mbox{\boldmath$\sigma$} 
 - \left[\mbox{\boldmath$n$}\times\left[\mbox{\boldmath$n$}
\times
\mbox{\boldmath$\xi$}_2\right]\right]\cdot\mbox{\boldmath$\sigma$},
\;\;\;\;\;\;\; \label{S}\\
\mbox{\boldmath$\xi$}_1 &=& \left(
(\alpha_1 p_{y'} + \alpha_2 x')/\lambda_-,\,
(\alpha_1 p_{x'} - \alpha_2 y')/\lambda_+,\, 0 \right),
\;\;\;\;\;\;\; \label{xi_1} \\
\mbox{\boldmath$\xi$}_2 &=& \left(
(\beta_1 p_{x'} + \beta_2 y')/\lambda_-,\,
(-\beta_1 p_{y'} + \beta_2 x')/\lambda_+,\, 0 \right),
\;\;\;\;\;\;\; \label{xi_2}\\
\alpha_1 &=& \frac{\hbar}{m^*}\frac{E_Z[E_Z^2 - (\hbar\omega_0)^2]}{(E_Z^2
- E_+^2)(E_Z^2- E_-^2)},\label{alpha1}\\
\alpha_2 &=& \frac{E_Z\hbar\omega_c(\hbar\omega_0)^2}{(E_Z^2
- E_+^2)(E_Z^2 - E_-^2)},\label{alpha2}\\
\beta_1 &=& \frac{\hbar}{m^*}\frac{E_Z^2  \hbar\omega_c}{(E_Z^2
- E_+^2)(E_Z^2
- E_-^2)},\label{beta1}\\
\beta_2 &=& E_Z^2\frac{(\hbar\omega_c)^2 + (\hbar\omega_0)^2 - E_Z^2}{(E_Z^2
- E_+^2)(E_Z^2 - E_-^2)},\label{beta2}
 \end{eqnarray}
where $ E_\pm = \hbar\omega \pm  \hbar\omega_c/2 $, with $\omega =
\sqrt{\omega_0^2 + \omega_c^2/4}$ and $\omega_c = eB_z/m^*c$.
Here, we assume $ E_{\pm} - |E_Z|  \gg  |E_Z\lambda_d/\lambda_{SO}|$,
which ensures that the lowest two levels in the quantum dot have spin nature.
Below, we consider low temperatures $T$
and bias $\Delta\mu$, such that $T,\Delta\mu \ll E_{\pm} - |E_Z| $, 
(hence only the orbital ground state is populated so that its Zeeman 
sublevels constitute a two level
system) and average over the dot ground
state in Eq. (\ref{Ht}). We  obtain, using Eqs. (\ref{L1})-(\ref{LZ}),
the following effective spin Hamiltonian
\begin{eqnarray}
H_{\rm eff} = \frac{1}{2} g \mu_B \left[{{\bm B}} + \delta
{{\bm B}}(t) \right ]
\cdot {{\bm \sigma}},\label{Heff}
\end{eqnarray} 
and the effective fluctuating magnetic field $\delta {{\bm B}}(t)$ 
 is then given by the operator
\begin{eqnarray}\label{dBfluct}
\delta {{\bm B}}(t) &=&
2{\bm B}  \times \left[{\bm \Omega}_1(t)+{\bm n}\times 
{\bm \Omega}_2(t)\right],\\
{\bm \Omega}_1 &=& \frac{e\hbar^2\gamma_1}{m^*} 
\left(\lambda_-^{-1}{\cal E}_{y'}, 
\lambda_+^{-1}{\cal E}_{x'}, 0\right),\nonumber\\
{\bm \Omega}_2 &=& \frac{e \hbar^2 \gamma_2}{m^*} 
\left(-\lambda_-^{-1}{\cal E}_{x'}, 
\lambda_+^{-1}{\cal E}_{y'}, 0\right),\nonumber\\
\gamma_1 &=& \frac{m^*}{\hbar E_Z} \alpha_1
= \frac{E_Z^2 - (\hbar\omega_0)^2}
{(E_+^2- E_Z^2)(E_-^2 - E_Z^2)},\nonumber\\
\gamma_2 &=& \frac{m^*}{\hbar E_Z} \beta_1
= \frac{E_Z  \hbar\omega_c}
{(E_+^2 - E_Z^2)(E_-^2 - E_Z^2)},\nonumber
\end{eqnarray}
 where we  have gone to the interaction picture with respect to the lead 
Hamiltonian $H_Q'= H_Q+  {\langle H_{Qd}\rangle}_d $  and
omitted a spin-independent part. 
Note that the coordinate-dependent part of $S$ drops out and thus
$\alpha_2$, $\beta_2$ do not enter.
Here and below, we use ${\langle\dots\rangle}_d$ to denote averaging 
over the dot ground state.  
Note that $H_Q'$ describes the QPC, while it is electrostatically 
influenced by the quantum dot with one electron in the ground state.
Obviously, $H_Q'$ can be rewritten in the same form as $H_Q$ 
in Eq. (\ref{HQ}), but with a different scattering phase in 
the scattering states. 
To denote the new scattering states,
we omit the {\em overbar} sign in our notations.
We have introduced an effective electric field operator 
$\mbox{\boldmath${\cal E}$}(t)$ in the interaction picture,
\cite{Mahan} 
\begin{eqnarray}
\mbox{\boldmath${\cal E}$}(t) &=&  \frac{1}{e} 
{\langle {{\bm \nabla}} H_{Qd}(t) \rangle}_d \nonumber\\
&=& \sum_{ll'kk'\sigma}\mbox{\boldmath${\varepsilon}$}_{ll'} 
 e^{i(\mu_l-\mu_{l'})t/\hbar} C^\dagger_{lk\sigma}(t)
 C_{l'k'\sigma}(t),\label{EQ}\\
\mbox{\boldmath${\varepsilon}$}_{ll'} &=&
\frac{1}{e} {\langle {{\bm \nabla}\eta_{ll'}({\bm r})} \rangle}_d,
\end{eqnarray} 
where  the fermionic operator $  C_{l'k'\sigma}$ 
corresponds to scattering states in
the leads with the dot being occupied by one electron ($H_Q'$ is
diagonal in $  C_{l'k'\sigma}$). Here, $\mu_l$, $l=L, R$, are the chemical 
potentials of the left ($L$) and right ($R$) leads, with 
$\Delta \mu =\mu_L-\mu_R$ being the voltage bias applied to the QPC
driving a current $I$. 
Note that in the absence of screening ($\tilde \delta({\bm R}-{\bm a}) =1$ 
in Eq.~(\ref{eta})), $\mbox{\boldmath${\cal E}$}$
coincides with the electric field 
that the quantum dot electron exerts on the QPC electrons.

 As a first result, we note that the fluctuating quantum field 
$\delta {{\bm B}}(t)$ is transverse with 
respect to the (classical) applied magnetic field ${{\bm B}}$
({\it cf}. Ref.~\onlinecite{GKL}).
The magnetic field fluctuations originate here from orbital
fluctuations that couple to the electron spin via the spin-orbit
interaction. The absence of time reversal symmetry, which is removed
by the Zeeman interaction, is crucial for this coupling.
We assume  no fluctuations in the external magnetic field 
${{\bm B}}$. In our model, the dot electron spin couples 
to a bath of fermions, in contrast to  Ref. \onlinecite{GKL} where the bath 
(given by phonons) was bosonic.

To calculate the coupling constants
$\mbox{\boldmath${\varepsilon}$}_{ll'}$ in Eq. (\ref{EQ})
, it is convenient to first
integrate over the coordinates of the dot electron. We thus obtain 
$\mbox{\boldmath${\cal E}$}({\bm R})=
\mbox{\boldmath${\cal E}$}_0({\bm R}) \tilde{\delta}({\bm R}-{\bm a})$,
 see Eq. (\ref{eta}), where ${\bm R}$ refers to the location of the electrons 
in the QPC and the bare (unscreened) electric field is given by
\begin{eqnarray} 
\mbox{\boldmath${\cal E}$}_0({\bm R}) &=& {e \over \kappa} 
\left\langle \frac{{\bm  R}-{\bm r}}{|{\bm  R}-{\bm r}|^3} 
\right\rangle_d \nonumber\\
&=& \frac{e {\bm R}}{\kappa R^3}
\left(1 + \frac{3}{4}\frac{\lambda_d^2}{R^2}+ \dots\right).
\label{ERQ}\;\;\;\;\;\;\;\;\;\;
\end{eqnarray}
Consequently, the coupling constants in Eq. (\ref{EQ}) read
$ \mbox{\boldmath${\varepsilon}$}_{ll'} = 
\langle lk|\mbox{\boldmath${\cal E}$}({\bm R})|l'k'\rangle $, where
$|lk\rangle$ denote the scattering states in the leads.
Here, we have assumed a parabolic confinement for the electron in the dot,
set the origin of coordinates in the dot center
 (${\langle {\bm r}\rangle}_d = 0 $) and averaged with the
dot wave  function $\Psi_d({\bm r}) =
 \exp {\left(- {r^2}/{2\lambda_d^2}\right)}/
 \lambda_d\sqrt{\pi}$, which is the ground state of the electron in 
 a symmetric harmonic potential in two dimensions.
While we choose a very special form for  the ground state wave
function, this does not affect substantially the final result, 
i.e. the relaxation time $T_1$. This is because any circularly symmetric
wave function leads to the same form for 
$\mbox{\boldmath${\cal E}$}_0({\bm R})$ except that it just alters the 
second term in Eq. (\ref{ERQ}) which is
very small compared to the first term (about one hundredth) and
negligible. An analogous argument applies to asymmetric 
wave functions. 

\section{Coupling Constants ${\bm \varepsilon}_{ll'}$}
\label{sec:CoupConst}
To proceed further, we construct the scattering states
out of the exact wave functions of an electron in the QPC potential.
While this is a generic method, we consider for
simplicity a $\delta$-potential tunnel barrier for the QPC,
\begin{equation}\label{VofX}
V(X) = \frac{\hbar^2 b}{m^*}\delta(X),
\end{equation}
where $b$ gives the strength of the delta potential.
Then, the electron wave functions in the even 
and odd channels are given by
\begin{eqnarray}  
\psi_e(X) &=&\sqrt{2}
\left\{\begin{array}{cc}
 \cos(kX + \phi),& X<0,\\
 \cos(kX - \phi),& X>0,
\end{array}\right.\label{cos}\\
\psi_o(X) &=& \sqrt{2}  \sin kX, \label{sin}
\end{eqnarray}
where $\phi = \arctan (b/k)$, $k=\sqrt{2m^*E/\hbar^2}$ and, for convenience,
the sample length is set to unity. Note that $\phi = \pi/2 - \delta$, where
$\delta \equiv \delta_e - \delta_o$ is the relative scattering phase
between the even ($e$) and odd ($o$) channels. 
The transmission coefficient
${\cal T}$ through the QPC is related to $\phi$ by 
${\cal T}(k)= \cos^2\phi$. We construct the scattering states in
the following way
\begin{eqnarray} 
\left(
\begin{array}{c}
\psi^L_{\rm sc}\\
\psi^R_{\rm sc}
\end{array}
\right) =
{\cal U}
\left(
\begin{array}{c}
\psi_e\\
\psi_o
\end{array}\right),
\;\;\;\;\;
{\cal U}={ -i \over {\sqrt 2} }\left(
\begin{array}{cc}
e^{i\delta}&-1\\
e^{i\delta}&1
\end{array}
\right).
\label{SC}
\end{eqnarray} 
Up to a global phase, Eq.~(\ref{SC}) is valid for any 
symmetric tunnel barrier. 

\subsection{Three limiting cases}\label{threelimcase}
We  calculate now the matrix elements of 
$\mbox{\boldmath${\cal E}$}({\bm R})$  using the wave
functions (\ref{cos}) and (\ref{sin}).
Three interesting regimes are studied in the following.

(i) $\lambda_{sc}\ll k_F^{-1} \ll a$, where $\lambda_{sc}$ is
the screening length in the QPC leads and $k_F$ is the Fermi wave vector. 
In this case, we set $\tilde \delta({\bm
 R} - {\bm a})= 2\lambda_{sc}\delta(X)$.
 By calculating the matrix elements of 
$\mbox{\boldmath${\varepsilon}$}$
with respect to the eigenstates of the potential barrier, Eqs. 
(\ref{cos}) and (\ref{sin}), we obtain 
\begin{eqnarray} 
\mbox{\boldmath${\varepsilon}$}_{ee} = 
4\lambda_{sc}
{\cal T}
\mbox{\boldmath${\cal E}$}_0({\bm a}),\label{epsee1}\;\;\;\;
\mbox{\boldmath${\varepsilon}$}_{oo}
=\mbox{\boldmath${\varepsilon}$}_{eo} = 0,
\end{eqnarray} 
where we used the odd and even eigenstates and 
$\int dY |\Phi(Y)|^2\mbox{\boldmath${\cal E}$}(X,Y)=
\mbox{\boldmath${\cal E}$}(X,a)$. Here,
$\Phi(Y)$ is  the QPC wave function in the  transverse direction
with width $\ll \lambda_{sc}$.
Going to the Left-Right basis, Eq. (\ref{SC}), 
which is more suitable for studying transport phenomena, we obtain 
\begin{eqnarray} 
\left(
\begin{array}{cc}
{\bm \varepsilon}_{LL}& {\bm \varepsilon}_{LR} \\
{\bm \varepsilon}_{RL}& {\bm \varepsilon}_{RR}
\end{array}
\right) &=&
{1 \over 2} {\bm \varepsilon}_{ee}
\left(
\begin{array}{cc}
1&1\\
1&1
\end{array}
\right).\label{epsll1} 
\end{eqnarray}
Note that in this case we have ${\bm \varepsilon}_{ll'} \propto 
{\cal T}$, where
 $l, l' = L, R$, see Eqs. (\ref{epsee1}) and (\ref{epsll1}).

(ii) $ k_F^{-1} \ll\lambda_{sc}\ll a$.
In this case, we  set $\tilde \delta({\bm
 R} - {\bm a})= \Theta (X + \lambda_{sc}) 
- \Theta (X - \lambda_{sc})$, where $\Theta (X)$ is the step function,
and we obtain in leading order in $1/k_F\lambda_{sc}$
\begin{eqnarray}
\mbox{\boldmath${\varepsilon}$}_{ee} &=&
\mbox{\boldmath${\varepsilon}$}_{oo} = 
\frac{2 e \lambda_{sc}}{\kappa  a^2}\left( 1 + {3\lambda_d^2 \over 4
  a^2} - {\lambda_{sc}^2 \over 2 a^2} + \dots
\right)\mbox{\boldmath${e}$}_Y,
\;\;\;\;\;\;\\
{\bm \varepsilon}_{eo} &=& \frac{ e  
\lambda_{sc}^2 \cos \delta}{\kappa  a^3}\left( 1 + {3\lambda_d^2 \over 4
  a^2} - {3\lambda_{sc}^2 \over 4 a^2} + \dots
\right)\mbox{\boldmath${e}$}_X.\label{epseo2}
\end{eqnarray}
In the above equations, $\mbox{\boldmath${e}$}_Y$ is a unit vector parallel to
${\bm a}$ and $\mbox{\boldmath${e}$}_X$ is a unit
vector perpendicular to ${\bm a}$ (see Fig. \ref{QPC}).
Further, we  assumed that $\hbar v_F \Delta k \le E_Z \ll 
 \hbar v_F \lambda_{sc}^{-1} \ll  E_F$, where $\Delta k = k-k'$,
$v_F$ is the Fermi velocity, and  $E_F = \hbar v_F k_F$ is the Fermi energy.
 Going as before to the Left-Right basis, we obtain
\begin{eqnarray}
\left(
\begin{array}{cc}
{\bm \varepsilon}_{LL}& {\bm \varepsilon}_{LR} \\
{\bm \varepsilon}_{RL}& {\bm \varepsilon}_{RR}
\end{array}
\right) &=&
\left(
\begin{array}{cc}
{\bm \varepsilon}_{ee}-{\bm \varepsilon}_{eo}\cos \delta&
i{\bm \varepsilon}_{eo}\sin \delta\\
-i{\bm \varepsilon}_{eo}\sin \delta&
{\bm \varepsilon}_{ee}+{\bm \varepsilon}_{eo}\cos \delta
\end{array}
\right).\;\;\;\;\;\;\;\;\label{epsll2} 
\end{eqnarray}
Note that in this case we have ${\bm \varepsilon}_{LR} \propto
\sqrt {{\cal T}(1-{\cal T})}$,  see Eqs.~(\ref{epseo2}) and (\ref{epsll2}). 
Since typically $\lambda_{sc} \gtrsim k_F^{-1}$, we expect case (ii) to describe
realistic setups. 
A more general case, $ k_F^{-1}, \lambda_{\rm sc} \ll a$, 
is studied in Appendix B.

(iii) $ k_F^{-1}, a \ll \lambda_{\rm sc}$.
In this regime, we neglect the screening 
($\tilde\delta ({\bm R}-{\bm a})=1$ in Eq.~(\ref{eta})). 
Then, we obtain the following 
expressions for the coupling constants
\widetext
\begin{eqnarray}
{\bm \varepsilon}_{oe} &=&  {\bm \varepsilon}_{eo} =
\frac{4ke}{\kappa}\left\{K_0(2ka)\sin \delta  
+ \frac{\pi}{2}\cos \delta \left[ I_0(2ka)-L_0(2ka) \right]\right\}
\label{case31}\mbox{\boldmath${e}$}_X,\\
{\bm \varepsilon}_{ee} &=&\frac{2e}{\kappa}
\left\{\frac{1}{a}-2k\cos(2\delta) K_1(2ka)
+\frac{\pi}{2}k\sin(2\delta)
\left[\frac{2}{\pi}-2I_1(2ka)+L_1(2ka)+L_{-1}(2ka)\right]\right\}
\mbox{\boldmath${e}$}_Y,\;\;\;\;\;\label{case32}\\
{\bm \varepsilon}_{oo}&=&\frac{2e}{\kappa}\left\{\frac{1}{a}-2k K_1(2ka)\right\}
\mbox{\boldmath${e}$}_Y,\label{case33}
\end{eqnarray}
\endwidetext
\noindent
where $I_n$ and $K_n$ are the modified Bessel functions and $L_n$ 
is the modified Struve function. Here, we assumed $ \Delta k \ll a^{-1} \ll 
\lambda_{sc}^{-1} $. 

Since usually $ka \gg 1$, the $k$-dependence of the coupling constants in 
Eqs.~(\ref{case31})-(\ref{case33}) is suppressed. 
One can use the following asymptotic expressions for $a \gg k_F^{-1}$, 
\begin{eqnarray}
{\bm \varepsilon}_{oe} &=&  {\bm \varepsilon}_{eo} \approx
\frac{2e\cos \delta}{\kappa a}
\mbox{\boldmath${e}$}_X,\label{case31a}\\
{\bm \varepsilon}_{ee} &\approx&{\bm \varepsilon}_{oo} \approx 
\frac{2e}{\kappa a}\mbox{\boldmath${e}$}_Y.\;\;\;\;\;\label{case32a}
\end{eqnarray}
In this case, the transformation to the Left-Right basis is given in 
Eq.~(\ref{epsll2}) and we obtain ${\bm \varepsilon}_{LR} \propto
\sqrt {{\cal T}(1-{\cal T})}$ as in case (ii).

\subsection{Consistency check}
Next we would like to verify whether our model predicts a realistic 
charge sensitivity of the QPC exploited in recent experiments.
\cite{Elzerman,Buks,Ensslin} 
For this we estimate the change in transmission $\delta \cal T$ 
through the QPC due to adding an electron to the quantum dot. 
The coupling in Eq.~(\ref{HQd}) (with coupling 
constants ${\eta_{ll'}({\bm r})}$ given in Eq.~(\ref{eta}))
is responsible for this transmission change $\delta \cal T$.
It is convenient to view this coupling as a potential $\delta V(X)$ induced
by the dot electron on the QPC. From Eq.~(\ref{eta}), we obtain 
\begin{eqnarray}
\delta V(X) &=& \frac{e^2}{\kappa\sqrt{X^2+a^2}}\tilde\delta(X),
\end{eqnarray}
where we have integrated over the dot coordinates
${\bm r}=(x,y)$ and the QPC coordinate $Y$, 
neglecting terms ${\cal O} (\lambda_d^2/a^2)$.
The screening factor $\tilde\delta(X)$ is peaked around $X=0$ with 
a halfwidth $\lambda_{sc}$. 
We consider two regimes.

(i) $\delta V(X)$ {\em is a smooth potential}. 
In this regime, $\hbar^2/m^*\bar a^2\ll\delta V(0)\ll E_F$, with 
$\bar a=\min (\lambda_{sc},a)$ being the width of $\delta V(X)$.
Therefore, the dot electron provides a constant potential 
(like a back gate) to the QPC, implying that 
$\delta V(X)$ merely shifts the origin of energy for the QPC electrons by
a constant amount, $\delta V(0)$.
From the geometry of the current experimental setups 
\cite{Elzerman,Buks,Ensslin} 
it appears reasonable to assume that this is the regime which 
is experimentally realized. The transmission change 
$\delta \cal T$ can then be estimated as
\begin{eqnarray}
&&\delta{\cal T} \approx -\delta V(0)
\left.\frac{\partial {\cal T}(E)}{\partial E}\right|_{E_F}\!\!\!\!
= -\frac{\delta V(0)}{E_F}{\cal T}(1-{\cal T}),\label{dT1}\;\;\;\;\;\\
&&{\cal T}(E)=\cos^2 \phi =\frac{E}{E+\hbar^2 b^2/2m^*},
\end{eqnarray}
where ${\cal T}={\cal T}(E=E_F)$.
By inserting typical numbers in Eq.~(\ref{dT1}), 
i.e. ${\cal T}=1/2$, $E_F=10\,{\rm meV}$, and $\delta V(0)=e^2/\kappa a$
$[\tilde\delta (0) =1]$, with $a=200\,{\rm nm}$ and $\kappa=13$, we obtain
$\delta{\cal T}/{\cal T}\approx 0.02$, 
which is consistent with the QPC charge sensitivity observed
experimentally.\cite{Elzerman}

(ii) $\delta V(X)$ {\em is a  sharp potential}. In this regime, 
adding an electron onto the quantum dot modifies the shape of the
existing tunnel barrier in the QPC.
Assuming sharp potentials, we obtain
\begin{eqnarray}
\delta \cal T &\approx& -\frac{2\delta A}
{A}{\cal T}(1-{\cal T}),\label{dT2}
\end{eqnarray}
where $\delta A=\int \delta V(X)dX$ and $A=\int V(X)dX=\hbar^2 b/m^*$.
In deriving Eq.~(\ref{dT2}), we assumed that $\delta A\ll A$.
Additionally, we assumed that both potentials $\delta V(X)$ and $V(X)$
are sharp enough to be replaced by $\delta$-potentials.
Redefining $\bar a$ such that $\delta A=\bar a \delta V(0)$,
we quantify the latter assumption as  $\bar a\ll 1/b$, where
$b$ is the strength of $V(X)$ in Eq.~(\ref{VofX}).
Note that for this regime the screening
is crucial, because $\delta  A \rightarrow \infty$ for 
$\lambda_{sc} \rightarrow \infty$.

\section{Spin Relaxation Time}
\subsection{$k$-independent case}
Next we use the effective Hamiltonian (\ref {Heff}) 
with Eqs.~(\ref{dBfluct}), (\ref{EQ}) and (\ref{epsll2}) to  calculate
the spin relaxation time $T_1$ of the electron spin on the dot
in lowest order in $\delta {\bm B}$. 
In the Born-Markov approximation, \cite{Slichter}
 the spin relaxation rate is given by \cite{GKL} 
$\Gamma_1 \equiv 1/T_1 = n_i n_j \Gamma_{ij}^r$,
where  ${\bm n}={\bm B}/B$ is the unit vector along
the applied magnetic field, $\Gamma_{ij}^r$ is
the spin relaxation tensor, and we imply summation 
over repeating indices.
To evaluate $T_1$, it is convenient to use the following expression,
obtained after regrouping terms in Ref.~\onlinecite{GKL},
\begin{eqnarray}
\frac{1}{T_1}&=&{\cal J}^+_{ii}(\omega_Z)-n_in_j{\cal J}^+_{ij}(\omega_Z)
-\epsilon_{kij}n_k{\cal J}^-_{ij}(\omega_Z),\;\;\;\;\;\;\label{Gamma}
\end{eqnarray}
where $\epsilon_{ijk}$ is the antisymmetric tensor 
and $\omega_Z =|E_Z|/\hbar$ is the Zeeman frequency. 
${\cal J}^\pm_{ij}(\omega_Z)$ are
Fourier transforms of anticommutators of the fluctuating fields
(with $\langle\delta {{\bm B}}(t)\rangle=0$)
\widetext
\begin{eqnarray}\label{calJp}
{\cal J}^+_{ij}(w)=\frac{g^2\mu_B^2}{4\hbar^2}\int_{-\infty}^{+\infty}
\langle\{\delta B_i(0), \delta B_j(t)\}\rangle \cos (w t) dt,\;\;\;\;\;\;\;\;
{\cal J}^-_{ij}(w)=\frac{g^2\mu_B^2}{4\hbar^2}\int_{-\infty}^{+\infty}
\langle\{\delta B_i(0), \delta B_j(t)\}\rangle \sin (w t)  dt,\;\;\;\;
\label{calJm}
\end{eqnarray}
\endwidetext
\noindent
which are evaluated in Eq. (\ref{Gamma}) at the Zeeman frequency
$\omega_Z$. Here and below, $\langle C \rangle \equiv 
{\rm Tr}(\rho_L\rho_R C)$ where $\rho_L$ ($\rho_R$) refers to the grand-canonical
density matrix of the left (right) lead at the chemical potential
$\mu_L$ ($\mu_R$), and ${\rm Tr}$ is the trace over the leads.
In our particular case, the second and third terms in
Eq.~(\ref{Gamma}) vanish.
The reason for vanishing of the second term is the
transverse nature of $\delta {{\bm B}}(t)$ in 
Eq.~(\ref{dBfluct}), {\em i.e.} $n_i\delta B_i(t)=0$.
The third term vanishes because each of the
$\bm{\varepsilon}_{ll'}$ in Eq. (\ref{epsll2})
is either real or imaginary.
The time dependence of the anticommutators of fluctuating 
fields at zero temperature, together with their Fourier transforms
(at finite temperature $T$) are
given by the following expressions
\begin{eqnarray}
\langle\{\delta B_i(0), \delta B_j(t)\}\rangle
&\propto& \frac{A(t)}{t^2},\\
{\cal J}^+_{ij}(w)&\propto& E_Z^2 {\cal S}(\hbar w),\;\;\;\;\;\;\;\;\;
\Delta \mu=0,\;\;\;\;\\
{\cal S}(x) &=& x \coth (x/2k_B T),
\end{eqnarray}
where $A(t)$ is an oscillatory function of $t$ with period $\Delta
\mu$ and ${\cal S}(\hbar w)$ is the spectral function of the 
QPC which is linear
in frequency at zero temperature. This time behavior shows that 
the QPC leads behave like an Ohmic bath.  
This Ohmic behavior results from bosonic-like particle-hole excitations
in the QPC leads, possessing a density of states that is linear 
in frequency close to the Fermi surface. In the spin-boson model, having
an Ohmic bath is sometimes problematic and  needs
careful study because of the non-Markovian effects of the bath.\cite{Spinboson}
However, we find that the Born-Markov approximation is still 
applicable since the non-Markovian corrections are not important in our
case, due to the smallness of the spin-orbit interaction. 
\footnote{In the spin-boson  model an appreciable non-Markovian
contribution emerges for coupling constants 
$\alpha =\hbar/T_1 E_Z \gtrsim 10^{-2}$. \cite{Spinboson} Since typically 
$\hbar/T_1 E_Z \lesssim 10^{-4}$ in the case we studied
 here [{\it cf}. Tables I, II],
we see that non-Markovian effects are negligible.}

For the fluctuating field $\delta {{\bm B}}(t)$,
we use the Born-Markov approximation\cite{Slichter}
and obtain from Eqs.~(\ref{Gamma}) and (\ref{calJm})
the spin relaxation rate
\begin{eqnarray}
\frac{1}{T_1}  &=& 
4\pi\hbar \nu^2 \left( M_{LL} + M_{RR} \right){\cal S}(E_Z) \nonumber\\ 
& & + 4\pi\hbar  \nu^2  M_{LR}\left[ {\cal S}(E_Z + \Delta\mu) + 
{\cal S}(E_Z - \Delta\mu) \right],\;\;\;\;\;\;\;\;
\label{relax}
\end{eqnarray}
where $\nu =1/2\pi\hbar v_F$ is the density of states
per spin and mode in the leads
  and the coefficients $M_{ll'}$ read
\begin{eqnarray}\label{Mllprime}
M_{ll'}&=&{\bm \omega}^{ll'}\cdot {\bm \omega}^{l'l}
-\left({\bm n} \cdot {\bm \omega}^{ll'}\right)   
\left({\bm n} \cdot {\bm \omega}^{l'l}\right),\\
{\bm \omega}^{ll'} &=& {\bm \Omega}^{ll'}_1 + {\bm n}\times 
{\bm \Omega}^{ll'}_2, \nonumber\\
{\bm \Omega}_1^{ll'} &=& \frac{e\hbar\gamma_1 E_Z}{m^*}
\left(\lambda_-^{-1}\varepsilon_{y'}^{ll'}, 
\lambda_+^{-1}\varepsilon_{x'}^{ll'}, 0\right),
\nonumber\\
{\bm \Omega}_2^{ll'} &=&\frac{e\hbar\gamma_2 E_Z}{m^*} 
\left(-\lambda_-^{-1}\varepsilon_{x'}^{ll'}, 
\lambda_+^{-1}\varepsilon_{y'}^{ll'}, 0\right), \nonumber
\end{eqnarray} 
where ${\bm \Omega}_i^{ll'}$ ($i=1,2$ and $l, l' = L, R$) 
are matrix elements of the
operators  ${\bm \Omega}_i$ with respect to the leads.
In addition, in deriving Eq. (\ref{relax}) we assumed 
$T, \Delta\mu \ll E_F$.
Note that, if the transmission coefficient of the QPC is  zero
or one (${\cal T} = 0, 1$), then Eq. (\ref{relax}) reduces to
\begin{eqnarray}
\frac{1}{T_1} =  4\pi\hbar \nu^2 (M_{LL}+ M_{RR}) E_Z,
 \;\;\;\;\;\;\;\;\;\;\;T \ll E_Z.
\end{eqnarray}
On the other hand, the equilibrium part of the relaxation 
time is obtained by assuming $\Delta\mu = 0$,
\begin{eqnarray}
\frac{1}{T_1} = 4 \pi\hbar  \nu^2  (
 M_{LL}+ M_{RR}+ 2M_{LR}) E_Z,\;\;\;\;\;T \ll E_Z.\;\;\;\;\;\label{T1eq}
\end{eqnarray}
Therefore, even with zero (or one) transmission coefficient 
or in the absence of the bias, the spin decay rate is non-zero due
to the equilibrium charge fluctuations in the leads.

\begin{table*}
\caption {Equilibrium ($\Delta \mu = 0$) relaxation time $T_1$ (ms) with 
${\bm B}$ along $x'$ (see Fig. \ref{QPC}).}
\begin{tabular}
{|c|c|c|c||c|c|}
\hline
 & & & & & \\
$T_1 \;(B=14\;\;T)$ & $T_1 \;(B=10\;\;T)$ & $T_1 \;(B=8\;\;T)$ 
& $T_1 \;(B=6\;\;T)$ &
 $\theta$ &  ${\cal T}$  \\
& & & & & \\
\hline \hline
0.9 & 2.77 & 5.64 & 13.78 & 0 & 0 \\
1.85 & 5.57 & 11.3 & 27.57  & 0 & 0.5 \\
$\infty$ & $\infty$ & $\infty$  & $\infty$  & 0 & 1 \\
0.1 & 0.32 & 0.66  & 1.62 & $\pi/4$ & 0\\ 
0.1 & 0.33 & 0.68 & 1.67 & $\pi/4$ & 0.5 \\
0.11 & 0.34 & 0.7  & 1.72 & $\pi/4$ & 1 \\ 
0.06 & 0.17 & 0.35 & 0.86 & $\pi/2$ & 0\\ 
0.06 & 0.17 & 0.35 & 0.86 & $\pi/2$ & 0.5 \\
0.06 & 0.17 & 0.35 & 0.86 & $\pi/2$ & 1 \\ 

\hline
\end{tabular}
\end{table*} 

\begin{table*}
\caption {Non-equilibrium ($E_Z \ll \Delta \mu =1$ meV) 
relaxation time $T_1$ (ms) with ${{\bm B}}$ along $x'$ (see Fig. \ref{QPC}).}
\begin{tabular}
{|c|c|c|c||c|c|}
\hline
& & & & & \\
$T_1 \;(B=14\;\;T)$ & $T_1 \;(B=10\;\;T)$ & $T_1 \;(B=8\;\;T)$ 
& $T_1 \;(B=6\;\;T)$ & $\theta$ &  ${\cal T}$  \\
& & & & & \\
\hline \hline
0.9 & 2.77 & 5.64 & 13.78 & 0  & 0 \\
0.95 & 2.25 & 3.8 & 7.32 & 0  & 0.5 \\
$\infty$ & $\infty$ & $\infty$   & $\infty$ & 0  & 1 \\
0.1 & 0.32 & 0.66 & 1.62 & $\pi /4$  & 0 \\
0.1 & 0.32 & 0.64 & 1.54 & $\pi /4$  & 0.5 \\
0.11 & 0.34 & 0.7 & 1.72 & $\pi /4$  & 1 \\
0.06 & 0.17 & 0.35 & 0.86 & $\pi /2$  & 0 \\
0.06 & 0.17 & 0.35 & 0.86 & $\pi /2$  & 0.5 \\
0.06 & 0.17 & 0.35 & 0.86 & $\pi /2$  & 1 \\
\hline

\end{tabular}
\end{table*} 

Another case of interest is the large bias regime 
$E_Z \ll \Delta\mu \ll \hbar \omega_0$,
 which simply means that only the second term in Eq. (\ref{relax})
appreciably contributes to the relaxation rate. Therefore, the 
non-equilibrium part of Eq. (\ref{relax}) is given by 
\begin{eqnarray}
\frac{1}{T_1}  \approx 
8\pi\hbar  \nu^2  M_{LR} \Delta\mu,\;\;\;\;\;
 E_Z, T \ll  | \Delta\mu \pm E_Z | \ll \hbar \omega_0.
\;\;\;\;\;\;\label{T1noneq}
\end{eqnarray}
To estimate the relaxation time, we use typical experimental
parameters for GaAs quantum dots (see, {\it e.g.}, 
Ref.~\onlinecite{ENature}). 
We consider an in-plane magnetic field ${\bm B}$ which leads to 
${\bm \Omega}_2=0$ ($\gamma_2=0$) and, 
for simplicity, assume that ${\bm B}$ is directed along one of the 
spin-orbit axes (say $x'$,  see Fig. \ref{QPC}). In this special case we obtain
the following expression for 
$k_F^{-1} \ll \lambda_{sc} \ll a$ (case (ii) of Sec.~\ref{threelimcase}), 
\begin{eqnarray}
M_{LR} \simeq \frac{e^4\hbar^2}
{{m^*}^2\kappa^2}
\frac{\lambda_{sc}^4}{\lambda_+^2 a^6}\frac{E_Z^2 \cos^2\theta}
{(\hbar^2\omega_0^2-E_Z^2)^2}
{{\cal T}(1-{\cal T})},
\end{eqnarray}
or equivalently, the relaxation rate is given in terms 
of the QPC shot noise
\begin{eqnarray}
\frac{1}{T_1}  &\approx&  \frac{8 \pi^2  e^2\hbar^4} {{m^*}^2\kappa^2}
\frac{\nu^2 \lambda_{sc}^4}{ a^6 \lambda_+^2}\frac{E_Z^2 \cos^2\theta}
{(\hbar^2\omega_0^2-E_Z^2)^2} S_{LL},\label{final}\\
S_{LL}&=& \frac{e^2 \Delta \mu}{\pi \hbar}{{\cal T}(1-{\cal T})}, 
\end{eqnarray}
where $S_{LL}$ is the current shot noise in the left lead of the QPC,
and due to current conservation,  $S_{LL} = S_{RR} = - S_{LR} 
=-S_{RL}$.\cite{Blanter}
We note that Eq. (\ref{final}) is the non-equilibrium part of the 
relaxation rate. Thus, even if the constant equilibrium part 
($\sim M_{LL},M_{RR}$ in Eq. (\ref{relax})) is of comparable
magnitude, the non-equilibrium part can  still be separated,  
owing to its bias dependence.
Moreover, at low temperatures and large bias voltages, the relaxation rate 
is linear in the bias $\Delta \mu$ and  proportional to the
current shot noise in the QPC, $1/T_1 \propto {\cal T}(1-{\cal T}) \Delta \mu$.
The latter relation holds in cases (ii) and (iii) of Sec.~\ref{threelimcase},
whereas in case (i) we have $1/T_1 \propto {\cal T}^2 \Delta \mu$.

The lifetime $T_1$ of the quantum dot spin strongly depends 
on the distance $a$ to the QPC. For the regime (ii) in Sec.~\ref{threelimcase},
the non-equilibrium part of $1/T_1$ depends on $a$ as follows,
$1/T_1 \propto a^{-6}$. 
A somewhat weaker dependence on $a$ occurs
in the regimes (i), $1/T_1 \propto a^{-4}$, and in the regime (iii), 
$1/T_1 \propto a^{-2}$.
On the other  hand, the charge sensitivity of the QPC scales as $a^{-1}$,
which allows one to tune the QPC into an optimal regime with reduced spin
decoherence but still sufficient charge sensitivity.

The spin lifetime $T_1$ strongly depends on the QPC orientation on the 
substrate (the angle $\theta$ between the axes $x'$ and $X$ 
in Fig.~\ref{QPC}). For example, in the regimes (ii) and (iii) 
(with $ka \gg 1$),
the non-equilibrium part of the relaxation rate vanishes at 
$\theta = \pi/2$, for an in-plane magnetic field ${\bm B}$ along $x'$.
Analogously, in the regime (i), both the equilibrium and the non-equilibrium
parts of the relaxation rate vanish at $\theta =0$, for 
${\bm B} \parallel x'$. 
 
We summarize our results in Tables I and II, where we have
evaluated the relaxation time $T_1$ (Eqs.~(\ref{T1eq}) and (\ref{relax}))
for a QPC located at $a=200$ nm away from the center of 
a GaAs quantum dot with $\lambda_d \approx 30$ nm, assuming
$\lambda_{sc} = 100$ nm, $\lambda_{SO} = 8\,\mu{\rm m}$,
and $k_F=10^8 \, {\rm m}^{-1}$. 
Here, we use coupling constants derived for the regime (ii) in
Sec.~\ref{threelimcase}.

Finally, we remark that, for a perpendicular magnetic field 
(${\bm B} =(0,0,B)$), we have 
\begin{eqnarray}
M_{ll'}=  {\bm \omega}^{ll'} \cdot {\bm \omega}^{l'l},\;\;\;\;\;\;\; 
\;\;\;\;\;\;\;{\bm n} ={\bm e}_z,
\end{eqnarray}
and the relaxation rate can be calculated analogously. The only
difference is that ${\bm \Omega}_2$ is no longer zero and the matrix
elements  $M_{ll'}$ are given by more complicated expressions.

\subsection{$k$-dependent case}
\begin{figure}[t]
 \begin{center}
  \includegraphics[angle=0,width=.45\textwidth]{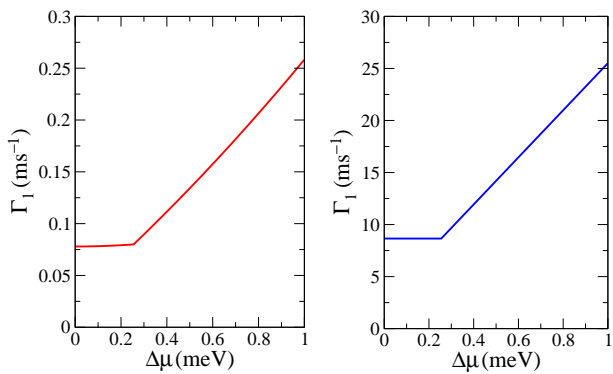}
    \caption{\small
Relaxation rate $\Gamma_1=1/T_1$ as a function of the bias $\Delta \mu$ 
applied to the QPC for cases (ii) and (iii), see Sec. IV.A. 
The magnetic field ${{\bm B}}$ is along $x'$ with magnitude $B=10$ T.
    }\label{L}
 \end{center}
\end{figure}

In this regime we use  the $k$-dependent coupling 
constants which are given in Eqs. (\ref{case31})-(\ref{case33}) 
and in Appendix B. Using Eq.~(\ref{Gamma}), the relaxation rate 
is given now by the following expression 
\widetext
\begin{eqnarray}
\frac{1}{T_1} &=& -\epsilon_{kij}n_k{\cal J}^-_{ij}(\omega_Z)
+ 4 \pi \hbar \nu^2 \sum_{ll'}\int dE \int dE' 
M_{ll'}(E,E')f(E)[1-f(E')] \nonumber\\
& & \times \{\delta(E' - E + \mu_{l'} - \mu_{l} - \hbar \omega_Z)+
\delta(E' - E + \mu_{l'} - \mu_{l} + \hbar \omega_Z)\},\label{T1k}
\end{eqnarray} 
\endwidetext
\noindent
where $f(E)=[\exp (E/k_BT)+1]^{-1}$ is 
the Fermi distribution function and  the 
energies are  measured from the Fermi level $\mu_l$ in each lead. 
The matrix elements $M_{ll'}(E,E')$ are given by Eq. (\ref{Mllprime}), 
however, in this case they are $k$-dependent through $E=\hbar v_Fk$.
Fig.~\ref{L} shows the numerical results for the relaxation rate $\Gamma_1=1/T_1$
as a function of the  bias $\Delta \mu$ for an in-plane magnetic 
field ${{\bm B}}$ of $10$ T in both cases. We note that
 the relaxation rate in case (iii) is typically two orders of magnitude
larger than in case (ii), which underlines the important role played by 
the screening length $\lambda_{sc}$ in the QPC-induced spin 
relaxation in a quantum dot.

\section{Concluding Remarks} 
In conclusion, we have shown that charge read-out devices 
(e.g. a  QPC charge detector) induces spin decay  in 
quantum dots due to the spin-orbit interaction (both Rashba and Dresselhaus).
Due to the transverse nature of the  fluctuating quantum
field $\delta {{\bm B}}(t)$, we found that pure dephasing is absent
and the spin decoherence time $T_2$ becomes twice the
relaxation time $T_1$, i.e. $T_2=2T_1$.
Finally, we showed  that the spin decay rate  is proportional to the shot noise
of the QPC in the regime of large bias ($\Delta\mu \gg E_Z$) and scales
as $a^{-6}$ (see Fig. \ref{QPC}). 
Moreover, we have shown that this rate can be minimized
by tuning certain geometrical parameters of the setup. 
Our results should  also be useful for designing experimental setups such
that the spin decoherence can be made negligibly small 
while charge detection with the QPC is still efficient. 
 
We thank J. Lehmann, W. A. Coish, T. Heikkil\"a, H. Gassmann and
S. Erlingsson  for helpful discussions.
This work was supported by the Swiss NSF, the NCCR
Nanoscience, EU RTN Spintronics, DARPA, and ONR.\\
\\

\appendix

\section{Schrieffer-Wolff transformation}
To derive the expression for $S$, we note that applying   
$\frac{1}{L_d^n}$ on ${\bm \xi}$  yields  linear combinations of
momentum and position operators. Therefore we make an ansatz for $S$, like
we did in Eq. (\ref{S}), with
\begin{eqnarray}
\mbox{\boldmath$\xi$}_1 &=& \left(
(\alpha_1 p_{y'} + \alpha_2 x')/\lambda_-,\,
(\tilde \alpha_1 p_{x'} + \tilde \alpha_2  y')/\lambda_+,\, 0 \right),
\;\;\;\;\;\;\; \\
\mbox{\boldmath$\xi$}_2 &=& \left(
(\beta_1 p_{x'} +\beta_2 y')/\lambda_-,\,
(\tilde \beta_1 p_{y'} + \tilde \beta_2 x')/\lambda_+,\, 0 \right).
\;\;\;\;\;\;\; 
\end{eqnarray}
Then by inserting this ansatz
into the relation $\left[ H_d + H_Z, S \right] =  H_{SO} $, we 
obtain a set of algebraic equations for the  coefficients 
$\alpha_i$, $\beta_i$, $\tilde \alpha_i$,
and $\tilde \beta_i$ ($i=1,2$). We find that
\begin{eqnarray}
&&\tilde \alpha_1 = \alpha_1,
\;\;\;\;\;\;\;\;
 \tilde \alpha_2 = - \alpha_2,\\
&&\tilde \beta_1 = -\beta_1, 
\;\;\;\;\;\;
\tilde \beta_2 =\beta_2,
\end{eqnarray}
with the  coefficients $\alpha_i$ and $\beta_i$
 given in Eqs. (\ref{alpha1})-(\ref{beta2}).

\section{$k$--dependent coupling constants, 
$ k_F^{-1} , \lambda_{\rm sc} \ll a$}

The coupling constants ${\bm \varepsilon}_{ee}$, 
${\bm \varepsilon}_{oo}$ and ${\bm \varepsilon}_{ee}$ 
are generally $k$-dependent. In the regime where 
$k_F^{-1} , \lambda_{\rm sc} \ll a$ we obtain the 
following relations 
\widetext
\begin{eqnarray}
{\bm \varepsilon}_{ee}
&=&\frac{e}{4 \kappa a^4 k^3}\{ 2 k^3 \lambda_{sc} (4 a^2 + 3 \lambda_d^2
- 2 \lambda_{sc}^2) + 6 k \lambda_{sc} \cos 2(k \lambda_{sc} + \delta) \nonumber\\
& &-(3 + 4 a^2 k^2 + 3 k^2 \lambda_d^2 - 6 k^2 \lambda_{sc}^2)
\sin 2(k \lambda_{sc} + \delta)  
+(3 + 4 a^2 k^2 + 3 k^2 \lambda_d^2) \sin (2\delta) \}
\mbox{\boldmath${e}$}_Y,\\
{\bm \varepsilon}_{oo}
&=&\frac{e}{4 \kappa a^4 k^3}\{ 2 k^3 \lambda_{sc} (4 a^2 + 3 \lambda_d^2
- 2 \lambda_{sc}^2) 
+ 6 k \lambda_{sc} \cos(2 k \lambda_{sc}) 
 - (3 + 4 a^2 k^2 + 3 k^2 \lambda_d^2 - 6 k^2 \lambda_{sc}^2)
\sin(2 k \lambda_{sc})\} \mbox{\boldmath${e}$}_Y,\\  
{\bm \varepsilon}_{oe}&=&
\frac{e}{8 \kappa a^5 k^4}\{ (9 + 4 a^2 k^2 + 3 k^2 \lambda_d^2 - 6 k^4
\lambda_{sc}^4 + 6 k^4 \lambda_d^2 \lambda_{sc}^2
+8 a^2 k^4 \lambda_{sc}^2)\cos \delta\nonumber\\ 
& &-(9 + 4 a^2 k^2 + 3 k^2 \lambda_d^2 - 18 k^2
\lambda_{sc}^2)\cos(2 k \lambda_{sc} + \delta)
-(9+4 a^2 k^2 + 3 k^2 \lambda_d^2 - 6 k^2 \lambda_{sc}^2)2 k 
\lambda_{sc}\sin(2 k \lambda_{sc} + \delta)\} \mbox{\boldmath${e}$}_X,
\;\;\;\;\;\;\;\;\;\;
\end{eqnarray}
\endwidetext
\noindent
with $\delta$ being the relative scattering phase.
The transformation to the Left-Right basis is given by
\begin{eqnarray}
{\bm \varepsilon}_{LL}&=&\frac{1}{2}(
{\bm \varepsilon}_{ee}+{\bm \varepsilon}_{oo}-
2{\bm \varepsilon}_{eo}\cos \delta),\\
{\bm \varepsilon}_{RR}&=&\frac{1}{2}(
{\bm \varepsilon}_{ee}+{\bm \varepsilon}_{oo}+
2{\bm \varepsilon}_{eo}\cos \delta),\\
{\bm \varepsilon}_{LR}&=&{\bm \varepsilon}_{RL}^*=\frac{1}{2}(
{\bm \varepsilon}_{ee}-{\bm \varepsilon}_{oo}+
2i{\bm \varepsilon}_{eo}\sin \delta).\label{AppElr}\;\;\;\;\;\;
\end{eqnarray}
Here, as before,
we have assumed that $\hbar v_F \Delta k \le E_Z \ll 
\hbar v_F \lambda_{sc}^{-1} \ll  E_F$.
Note that the coupling constants  ${\bm \varepsilon}_{LR}$ and
${\bm \varepsilon}_{RL}$ in Eq.~(\ref{AppElr}) have both real and 
imaginary parts. Therefore, the last term in Eq.~(\ref{Gamma}) 
does not vanish in general.
Nevertheless, we find that for an in-plane magnetic field 
${\bm B}=(B_x,B_y,0)$ this term vanishes, because only a single component
of $\delta {\bm B}(t)$ (namely $\delta B_z(t)$, see Eq.~(\ref{dBfluct}))
 is present for in-plane fields, which leads to
$\epsilon_{kij}n_k{\cal J}^-_{ij}(\omega_Z)=0$ 
(see also Eqs.~(\ref{calJm}) and (\ref{T1k})).

\end{document}